\def\mumu{$\mu^+\mu^-$}
\def\ee{$e^+e^-$}

\documentstyle[epsfig,longtable]{aipproc}

\begin{document}
\title{Muon Colliders: New Prospects for Precision Physics and the High
Energy Frontier}
\author{Bruce J. King$^1$}
\address{Brookhaven National Laboratory\\
email: bking@bnl.gov\\
web page: http://pubweb.bnl.gov/people/bking/}
\thanks{
Presented at the Latin American Symposium on High Energy Physics,
April 8-11, 1998, San Juan, Puerto Rico.
%..This updated version from
%..September 9, 1998 includes corrections to the values given in table 1
%..for the predicted number of events in a long baseline neutrino
%..detector.
This work was performed under the auspices of
the U.S. Department of Energy under contract no. DE-AC02-98CH10886.}
\maketitle

\begin{abstract}
   An overview is given of muon collider technology and of the
current status of the muon collider research program. The exciting
potential of muon colliders for both neutrino physics and collider physics
studies is then
described and illustrated using self-consistent collider parameter sets
at 0.1 TeV to 100 TeV center-of-mass energies.
\end{abstract}

\section*{Introduction}
%%%%%%%%%%%%%%%%%%%%%%%
\label{sec-intro}

%\subsection*{Motivation for a New Accelerator Technology}
%\label{subsec-motive}

 Muon colliders appear to be emerging as a promising complement and/or
alternative to proton and electron colliders for experimental high energy
physics (HEP) studies at the high energy frontier. They also provide
some interesting possibilities for precision studies in HEP, particularly
in neutrino physics.

 This paper consists of three main sections. The first section gives a
very brief description of muon collider technology then two
longer sections give introductions to the neutrino physics potential and
collider physics potential of muon colliders, respectively.

%\newpage

\squeezetable
\begin{table}[htb!]
%\begin{center}
\caption{
Self-consistent parameter sets for muon colliders at CoM energies ranging from
0.1 TeV to 100 TeV. For completeness, beam parameters and collider ring
parameters have been included along with the physics parameters, and
the generation and optimization of these parameter sets is described
in reference [2]. Except for the first parameter set, which has been studied
in some detail by the Muon Collider Collaboration, these parameters represent
speculation by the author on how muon colliders might evolve with energy.}
\begin{tabular}{|r|ccccc|}
\hline
\multicolumn{1}{|c|}{ {\bf center of mass energy, ${\rm E_{CoM}}$} }
                            & 0.1 TeV & 1 TeV & 4 TeV &  10 TeV  & 100 TeV \\
\multicolumn{1}{|c|}{ {\bf description} }
                            & Higgs factory & LHC complement & E frontier &
                                        $2^{\rm nd}$ gen. & ult. E scale \\
\hline \hline
\multicolumn{1}{|l|}{\bf collider physics parameters:} & & & &  & \\
luminosity, ${\cal L}$ [${\rm cm^{-2}.s^{-1}}$]
                                        & $1.0 \times 10^{31}$
                                        & $1.0 \times 10^{34}$
                                        & $6.2 \times 10^{33}$
                                        & $1.0 \times 10^{36}$
                                        & $4.0 \times 10^{36}$ \\
$\int {\cal L}$dt [${\rm fb^{-1}/det/year}$]
                                        & 0.1 & 100 & 62 & 10 000
                                        & 40 000 \\
No. of $\mu\mu \rightarrow {\rm ee}$ events/det/year
                                        & 870 & 8700 & 340 & 8700 & 350 \\
No. of 100 GeV SM Higgs/det/year        & 3700 & 69 000
                                         & 69 000 & $1.4 \times 10^7$
                                         & $8.3 \times 10^7$ \\
fract. CoM energy spread, ${\rm \sigma_E/E}$ [$10^{-3}$]
                                        & 0.02 & 1.6 & 1.6 & 1.0 & 1.0 \\
\hline
\multicolumn{1}{|l|}{\bf neutrino physics parameters:}   & & & & & \\
fract. str. sect. length, ${\rm f_{ss}}$  & 0.15 & 0.10 & 0.05 & 0.04 & 0.02\\
neutrino ang. divergence, $\theta_\nu$[$1/\gamma$]
                                           & 1 & 10 & 10 & 10 & 10 \\
high rate det: events/yr/(${\rm g.cm^{-2}}$) & $8.1 \times 10^6$
                                             & $1.9 \times 10^7$
                                             & $1.5 \times 10^6$
                                             & $1.3 \times 10^8$
                                             & $2.5 \times 10^7$ \\
long baseline: events/yr/(${\rm kg.km^{-2}}$) & $1.8 \times 10^5$
                                              & $4.2 \times 10^5$
                                              & $5.3 \times 10^5$
                                              & $2.9 \times 10^8$
                                              & $5.6 \times 10^9$ \\
\hline
\multicolumn{1}{|l|}{\bf collider ring parameters:}   & & & & & \\
circumference, C [km]                   & 0.3 & 2.0 & 7.0 & 15 & 100 \\
ave. bending B field [T]               & 3.5 & 5.2 & 6.0 & 7.0 & 10.5 \\
\hline
\multicolumn{1}{|l|}{\bf beam parameters:}            & & & & & \\
($\mu^-$ or) $\mu^+$/bunch,${\rm N_0[10^{12}}]$
                                        & 4.0 & 3.5 & 3.1 & 2.4 & 0.18 \\
($\mu^-$ or) $\mu^+$ bunch rep. rate, ${\rm f_b}$ [Hz]
                                        & 15 & 15 & 0.67 & 15 & 60 \\
6-dim. norm. emittance, $\epsilon_{6N}
               [10^{-12}{\rm m}^3$]    & 170 & 170 & 170 & 50 & 2 \\
x,y emit. (unnorm.)
              [${\rm \pi.\mu m.mrad}$] & 710 & 12 & 3.0 & 0.55 & 0.0041 \\
x,y normalized emit.
              [${\rm \pi.mm.mrad}$]    & 340 & 57 & 57 & 26 & 1.9 \\
fract. mom. spread, $\delta$ [$10^{-3}$]
                                       & 0.03 & 2.3 & 2.3 & 1.4 & 1.4 \\
relativistic $\gamma$ factor, ${\rm E_\mu/m_\mu}$
                                        & 473 & 4732 & 18 929 & 47 322
                                        & 473 220 \\
ave. current [mA]                      & 20 & 10 & 0.46 & 24 & 4.2 \\
beam power [MW]                        & 1.0 & 8.4 & 1.3 & 58 & 170 \\
decay power into magnet liner [kW/m]   & 1.1 & 0.58 & 0.03 & 1.4 & 1.3 \\
time to beam dump,
          ${\rm t_D} [\gamma \tau_\mu]$ & no dump & 0.5 & 0.5 & no dump & 0.5 \\
effective turns/bunch                  & 519 & 493 & 563 & 1039 & 985 \\
\hline
\multicolumn{1}{|l|}{\bf interaction point parameters:}      & & & & & \\
spot size, $\sigma_x = \sigma_y
                          [\mu {\rm m}]$   & 270 & 7.6 & 1.9 & 0.78 & 0.057 \\
bunch length, $\sigma_z$ [mm]          & 11 & 4.7 & 1.2 & 1.1 & 0.79 \\
$\beta^*$ [mm]                          & 11 & 4.7 & 1.2 & 1.1 & 0.79 \\
ang. divergence, $\sigma_\theta$
                             [mrad]    & 2.6 & 1.6 & 1.6 & 0.71 & 0.072 \\
beam-beam tune disruption parameter, $\Delta \nu$
                                        & 0.013 & 0.066 & 0.059 & 0.100
                                        & 0.100 \\
pinch enhancement factor, ${\rm H_B}$  & 1.000 & 1.040 & 1.025 & 1.108
                                        & 1.134 \\
beamstrahlung fract. E loss/collision  & $5 \times 10^{-16}$
                                       & $1.2 \times 10^{-10}$
                                       & $2.3 \times 10^{-8}$
                                       & $2.3 \times 10^{-7}$
                                       & $3.2 \times 10^{-6}$ \\
\hline
\multicolumn{1}{|l|}{\bf final focus lattice parameters:} & & & & & \\
max. poletip field of quads., ${\rm B_{4\sigma}}$ [T]
                                        & 6 & 10 & 10 & 15 & 20 \\
max. full aperture of quad., ${\rm A_{\pm4\sigma}}$[cm]
                                        & 14 & 13 & 30 & 20 & 13 \\
${\rm \beta_{max} [km]}$               & 0.4 & 22 & 450 & 1100
                                        & 61 000 \\
final focus demagnification, $\sqrt{\beta_{\rm max}/\beta^*}$
                                        & 60 & 2200 & 19 000 & 31 000
                                       & 280 000 \\
\hline
\multicolumn{1}{|l|}{\bf synchrotron radiation parameters:} & & & & & \\
syn. E loss/turn [MeV]                 & 0.0008
                                        & 0.01 & 0.9 & 17 & 25 000 \\
syn. rad. power [kW]                   & 0.0002 & 0.13 & 0.4 & 400
                                        & 110 000 \\
syn. critical E [keV]                  & 0.0006 & 0.09 & 1.6 & 12 & 1700 \\
\hline
\multicolumn{1}{|l|}{\bf neutrino radiation parameters:} & & & & & \\
collider reference depth, D[m]           & 10 & 125 & 300 & 300 & 300 \\
ave. rad. dose in plane [mSv/yr]         & $3 \times 10^{-5}$
                                        & $9 \times 10^{-4}$
                                        & $9 \times 10^{-4}$
                                        & 0.66 & 6.7 \\
str. sect. length for 10x ave. rad.,
                     ${\rm L_{x10}}$[m] & 1.9 & 1.3 & 1.1 & 1.0 & 2.4 \\
$\nu$ beam distance to surface [km]    & 11 & 40 & 62 & 62 & 62 \\
$\nu$ beam radius at surface [m]       & 24 & 8.4 & 3.3 & 1.3 & 0.13 %\\ \hline

\end{tabular}
\label{specs}
%\end{center}
\end{table}

%\newpage

 The two physics sections use, as examples, the muon
collider parameter sets of table~\ref{specs}, at center of mass (CoM)
energies from 0.1 TeV to 100 TeV. The parameter set at 0.1 TeV CoM energy,
which is intended as an s-channel
Higgs factory, was constrained to essentially reproduce one of the
parameter sets currently under study~\cite{status} by the Muon Collider
Collaboration (MCC). In contrast, the other sets represent
speculation by the author on how the parameters might evolve with CoM energy.
A discussion and assessment of the technical challenges associated with
these specific parameter sets is given
in~\cite{epac98}. It should be stressed that they are all still
rather speculative (additional to the rather immature status
of the entire muon collider technology) and have not been
studied or discussed in detail within the MCC. This applies even more
strongly to the final parameter set, at 100 TeV, which might represent
the ultimate energy scale for muon colliders and which assumes technological
extrapolations (in magnets, etc.) that might not come to pass for
at least another couple of decades.

\section*{An Overview of Muon Colliders}
%%%%%%%%%%%%%%%%%%%%%%%%%%%%%%%%%%%%%%%%
\label{sec-overview}

%\subsection*{History and Current Status of Studies}
%\label{subsec-status}

  The technology of muon colliders is relatively new. The
possibility of muon colliders was introduced by Budker~\cite{budker},
Skrinsky et al.~\cite{ref2} and
Neuffer~\cite{ref3} and has been aggressively developed over the past four
years in a series of collaboration meetings and
workshops~\cite{ref4,ref5,ref6,ref7}. A detailed feasibility study for a
4 TeV muon collider~\cite{book} was presented at Snowmass96
and, since then, progress has continued on studies for both
this collider and others at lower energies. The current status of
MCC studies is summarized in~\cite{status}.
The Muon Collider Collaboration now consists
of over 100 physicists and engineers from the U.S.A., Europe and Japan,
largely based in the U.S.A. and mainly at three U.S. national laboratories:
Brookhaven National Laboratory (BNL), Fermi National
Accelerator Laboratory (FNAL) and Lawrence Berkeley National Laboratory
(LBNL).

%\subsection*{An Overview of the Components}
%\label{subsec-comp}

%...*** figure with muon collider components ***
\begin{figure}[t!] % fig 1
\centering
\includegraphics[height=2.5in,width=5.0in]{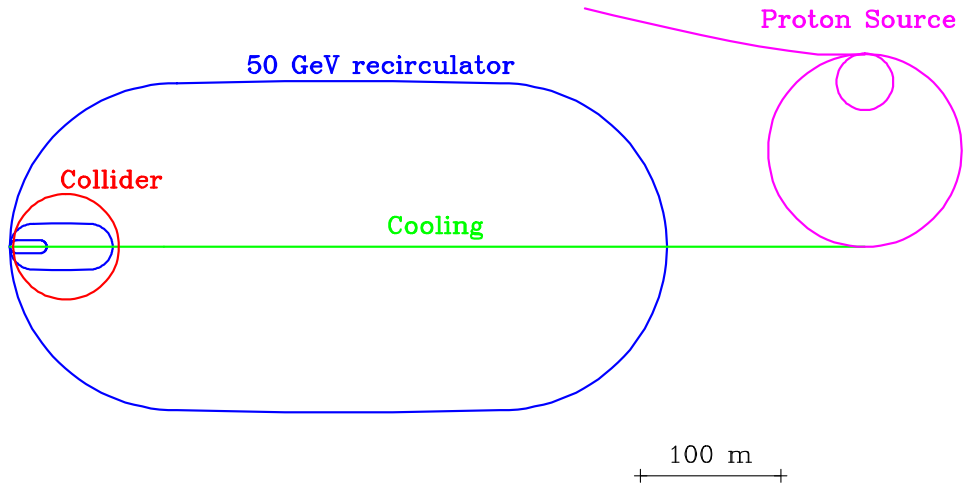}
\caption{Schematic footprint of a 100 GeV muon collider (reproduced
from reference [1]).}
\label{schematic}
\end{figure}

%  The basic components of the \mumu collider are shown schematically in 
%Fig.\ref{schematic}. The muons travel through the components sequentially
%from top to bottom of the figure.

  Figure~\ref{schematic} illustrates the basic layout of a \mumu collider
using, as an example, the schematic footprint of a 0.1 TeV collider.
Initially, large bunches of low energy muons are produced by targeting
proton bunches from a high intensity proton source onto a pion production
target inside a solenoidal capture and decay channel. The relatively diffuse
muon bunches from the decay channel then enter an ionization cooling channel,
which shrinks them down to a suitable emittance for fast acceleration and
injection, at full energy, into a collider storage ring. (The final
acceleration stage is anomalously larger than the collider ring for
the low energy collider of figure~\ref{schematic}. Depending on design
and technology choices, the final acceleration stage may well remain larger
than the collider ring for higher energy colliders, but likely by a lesser
margin.)

  The ionization cooling channel is the most novel and characteristic feature
of a muon collider, and also the biggest technical challenge.
As a general outline of the cooling process, the muons in
each bunch lose both transverse and longitudinal momentum in passing through
a material medium and are then reaccelerated in radiofrequency (rf) cavities,
restoring the
longitudinal momentum but leaving a reduced transverse momentum spread in
the bunch. Also, the momentum spread of the bunch can be reduced by using
wedges of material in a dispersive section of a magnet lattice to reduce
preferentially the momenta of the munos with higher momenta. A
large amount of cooling is required -- current scenarios give a factor of
$10^6$ reduction in the invariant 6-dimensional phase space -- so the
cooling channel
will probably be a repetitive structure with perhaps 20 to 30 stages.
The MCC is pursuing a vigorous theoretical
and experimental program to develop and test the components of the
cooling channel.

   Because of the short muon lifetime -- 2.2 microseconds in the muon rest
frame -- the muon cooling and acceleration must be done very quickly.
Current scenarios envisage about a 50\% decay loss in the cooling channel
and a 25\% loss of the remaining muons during acceleration.
Also, the muons survive for only of order 1000 turns in the collider ring
(almost independent of the collider energy), so the muon bunches
must be frequently replenished.
Undesirable consequences of the large bunches of muons decaying to
electrons are the resulting large and difficult background in the
collider detectors, the radiation heat load on the collider ring and,
surprisingly, the potential radiation hazard from the intense neutrino
beams. The neutrino radiation hazard becomes an important design
constraint for high energy colliders, and is discussed in more detail
in a later section.

%
%The number of muons per bunch remaining ($N_f$) after a stage of acceleration
%in a recirculating ring can be calculated from the initial number ($N_i$)
%using the simple, easily derived formula:
%\begin{equation}
%ln(N_f/N_i) = 0.16 ln(E_i/E_f) \times L[km] / E_rf[GeV] ,
%\end{equation}
%where
%$E_i, E_f =$ initial, final energies,
%L = effective circumference,
%$E_rf =$ rf acceleration per turn.
%

\section*{Prospects for Neutrino Physics}
%%%%%%%%%%%%%%%%%%%%%%%%%%%%%%%%%%%%%%%%%
\label{sec-nuphys}

  This section gives an overview of the neutrino physics
possibilities at a future muon storage ring, which can be either a muon
collider ring or a ring dedicated to neutrino physics that uses muon
collider technology to store large muon currents.
It summarizes a previous more detailed description of these topics
by this author~\cite{nufnal97} (using a generalized description of neutrino
production and event rates that is now applicable to all muon colliders).

  The section begins with a characterization of the neutrino beam
and predictions for neutrino event rates in both general purpose and
long-baseline neutrino detectors, then follows with a description of a
specific design for a general baseline detector. Finally,
an overview is given of some of the important physics analyses that
could be performed at such ``muon ring neutrino experiments'' (MURINE's).

\subsection*{Neutrino Beam and Experimental Overview}
\label{subsec-nuprod}

  Neutrinos are emitted from the decay of muons in the collider ring:
\begin{eqnarray}
\mu^- & \rightarrow & \nu_\mu + \overline{\nu_{\rm e}} + {\rm e}^-,
                                             \nonumber \\
\mu^+ & \rightarrow & \overline{\nu_\mu} + \nu_{\rm e} + {\rm e}^+.
                                                 \label{eq:nuprod}
\end{eqnarray}

  The thin pencil beams of neutrinos for experiments will be produced
from long straight sections in either the collider ring or a
ring dedicated to neutrino physics.
From relativistic kinematics, the
forward hemisphere in the muon rest frame will be boosted, in the lab
frame, into a narrow cone with a characteristic opening half-angle,
$\theta_\nu$,
given in obvious notation by
\begin{equation}
\theta_\nu \simeq \sin \theta_\nu = 1/\gamma =
\frac{m_\mu}{E_\mu} \simeq \frac{10^{-4}}{E_\mu ({\rm TeV})}.
                                                   \label{eq:thetanu}
\end{equation}
The final focus regions around collider
experiments are important exceptions to equation~\ref{eq:thetanu} since
the muon beam itself will have an angular divergence in these regions that
is large enough to spread out the neutrino beam by at least an order of
magnitude in both x and y. It is likely that neutrino experiments
at sub-TeV CoM energy muon colliders will use the beams from either dedicated
or utility straight sections opposite the collider detector while those
at higher energy muon colliders -- where neutrino radiation is an
important design constraint -- will use the more divergent beam emanating
from the final
focus region. A dedicated storage ring could avoid the problem
of neutrino radiation by using a long downward-tilting long straight section.

 The dominant interaction of TeV-scale neutrinos is deep
inelastic scattering (DIS) off nucleons with the production of several hadrons.
This is reinterpreted in the quark-parton model as elastic or quasi-elastic
scattering off the quark constituents of the nucleons followed by
hadronization of the final state quark. Charged current (CC) DIS scattering,
which is mediated by a charged W boson and comprises about 75\% of the total
cross-section, may be represented as
\begin{eqnarray}
\nu + q & \rightarrow & l^- + q', \nonumber \\
\overline{\nu} + q' & \rightarrow & l^+ + q
                                        \label{eq:ccq},
\end{eqnarray}
where $l$ is an electron/muon for electron/muon neutrinos and the quarks,
($q$) and ($q'$), have charges differing by one unit.
Neutral current (NC) DIS scattering,
\begin{equation}
\nu + q \rightarrow \nu + q,
                                        \label{eq:ncq}
\end{equation}
 which is interpreted as neutrino-quark
elastic scattering with the exchange of a neutral
Z boson, makes up the remaining 25\% of the cross-section.

%..# interactions in a detector (2 equations)

  For TeV-scale neutrinos, the neutrino cross-section is approximately
proportional to the neutrino
energy, $E_\nu$, and the charged current (CC) and
neutral current (NC) interaction cross sections for neutrinos and
antineutrinos have numerical values of~\cite{quigg}:
\begin{equation}
 {\rm \sigma_{\nu N}\; for\;}
 \left(
 \begin{array}{c}
   \nu-CC \\
   \nu-NC \\
   \overline{\nu}-CC \\
   \overline{\nu}-NC
 \end{array}
 \right)\;
 \simeq
 \left(
 \begin{array}{c}
    0.72 \\ 0.23 \\ 0.38 \\ 0.13
   \end{array}
 \right)
\times {\rm E_\nu [TeV]}
\times 10^{-35}\: {\rm cm^2}.
                                            \label{eq:xsec}
\end{equation}

  Using these cross-section values, it is straightforward to derive
predictions for the approximate neutrino event rates at a neutrino detector.
For a general purpose detector subtending the boosted forward hemisphere
of the neutrino beam:
\begin{eqnarray}
{\rm Number\; of\; \nu\; events/yr} & \simeq &
          1.8 \times 10^7 \times l [{\rm g.cm^{-2}}] \nonumber \\
  & & \times {\rm f_b[Hz] \times N_0[10^{12}] \times E_\mu [TeV] \times f_{ss}
                   \times (1-e^{-t_D[\gamma \tau_\mu]}) },
                                            \label{eq:genevents}
\end{eqnarray}
with notation as in table~\ref{specs},
where $l$ is the detector length, $10^7$ seconds of running time
per year are assumed and the fractional breakdown into interaction
types is as in equation~\ref{eq:xsec}.

  The analagous equation for
a long baseline detector in the center of the neutrino beam is:
\begin{eqnarray}
{\rm Number\; of\; \nu\; events/yr} & \simeq &
          9 \times 10^6 \times {\rm \frac{M[kg]}
       {(L[km])^2 \times ( \gamma \theta_\nu)^2} } \nonumber \\
  & & \times {\rm f_b[Hz] \times N_0[10^{12}] \times E_\mu [TeV] \times f_{ss}
                   \times (1-e^{-t_D[\gamma \tau_\mu]})  },
                                            \label{eq:longevents}
\end{eqnarray}
where $M$ is the detector mass, $L$ the distance from the neutrino
source and the factor $( \gamma \theta_\nu )^{-2}$ allows for the
possibility that the divergence, $\theta_\nu$, of the neutrino beam
is larger than $1/\gamma$. Using these equations, table 1 gives
numerical predictions of event rates for each of the
parameter sets. Clearly, these can be several orders of magnitude
higher than at today's neutrino beams, even when using less massive
targets.

\subsection*{A General Purpose Neutrino Detector}
\label{subsec-nudet}

%..detector figure
%..table of subdetectors
%..overview of subdetectors

%...*** neutrino detector figure ***
\begin{figure}[t!] %
\centering
\includegraphics[height=3.5in,width=3.5in]{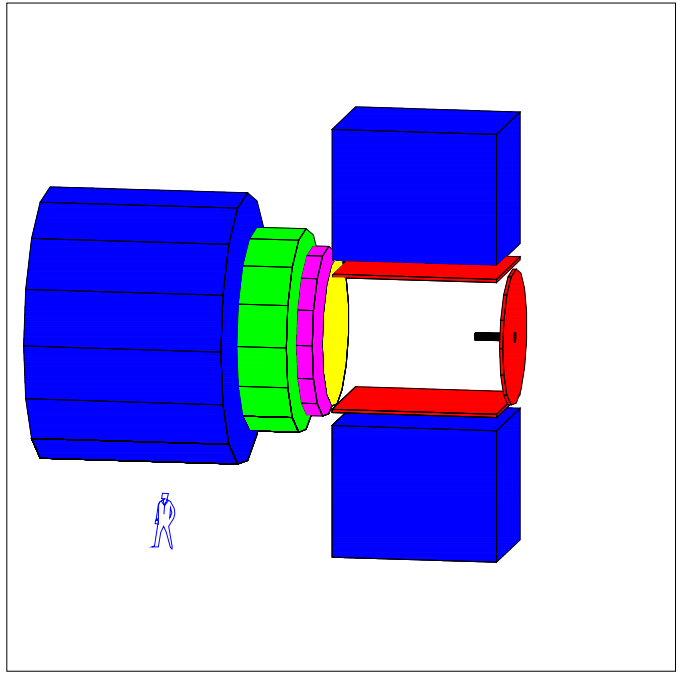}
\caption{Example of a general purpose neutrino detector,
reproduced from reference [12]. A human figure in
the lower left corner illustrates its size. The neutrino target is the small
horizontal cylinder at mid-height on the right hand side of the detector. Its
radial extent corresponds roughly to the radial spread of the neutrino pencil
beam, which is incident from the right hand side. Further details are given
in the text.}
\label{detector_fig}
\end{figure}

Figure~\ref{detector_fig} is an example of the sort of high rate general
purpose neutrino detector that would be well matched to the intense neutrino
beams at muon colliders.
The neutrino target is a 1 meter long stack of CCD tracking planes with
a radius of 10 cm chosen to
match the beam radius at approximately 200 meters from
production for a 250 GeV muon beam.
 It contains 750 planes of 300 micron thick silicon CCD's, corresponding to
a mass per unit area of approximately 50 ${\rm g.cm^{-2}}$, about
2.5 radiation lengths and 0.5 interaction lengths.
(Note the contrast with the kilotonne-scale
calorimetric targets used in today's high rate neutrino experiments.)
For this target, it is seen that the parameter sets in table 1 typically
correspond to  several hundred million neutrino interactions per year,
and the rate could be even higher for a dedicated muon storage ring or
more massive target.

  Besides providing the mass for neutrino
interactions, the tracking target allows precise reconstruction of the event
topologies from charged tracks, including event-by-event vertex tagging
of those events containing charm or beauty hadrons or tau
leptons. Given the favorable vertexing geometry and the few-micron typical
CCD hit resolutions, it is reasonable to expect almost 100 percent
efficiency for b tagging, perhaps 70 to 90 percent efficiency for
charm tagging and excellent discrimination between b and c decays.

 The target in figure~\ref{detector_fig} is surrounded by a
time projection chamber (TPC) tracker in a vertical
dipole magnetic field. The
characteristic dE/dx signatures from the tracks would identify
each charged particle. Further particle ID is provided by the Cherenkov
photons that are produced in the TPC gas then reflected by a spherical
mirror at the downstream end of the tracker and focused onto a read-out
plane at the upstream end of the target.
The mirror is backed by electromagnetic and hadronic calorimeters
and, lastly, by iron-core toroidal magnets for muon ID.

 The relativistically invariant quantities that are routinely extracted in
DIS experiments are 1) Feynman x, the fraction of the nucleon momentum
carried by the struck quark, 2) the inelasticity,
${\rm y = E_{hadronic}/E_\nu}$,
which is related to the scattering angle of the neutrino in the neutrino-quark
CoM frame, and 3) the momentum-transfer-squared,
${\rm Q^2 = 2 M_{proton} E_\nu x y}$. As a significant advance,
this detector will have the further capability
of accurately reconstructing the hadronic 4-vector, resulting in a much better
characterization of each interaction, particularly for NC interactions.

 Another big improvement over today's neutrino detectors is the vastly
improved ability to reconstruct the flavor of the final state quark.
Final state c and b quarks can be identified by vertex tagging
of the decaying charm or beauty hadrons that contain them, and
some statistically based flavor tagging will also be available for u, d or s
final state quarks, taking advantage of the ``leading particle effect''
that is used, for example, in LEP analyses of hadronic Z decays.

\subsection*{Neutrino Physics Opportunities}
%%%%%%%%%%%%%%%%%%%%%%%%%%%%%%%%%%%%%%%%%%%%

 Neutrino interactions are interesting both in their own right and as probes
of the quark content of nucleons, so a MURINE has wide-ranging potential to
make advances in many areas of elementary particle physics. This section gives
an overview for measurements involving the Cabbibo-Kobayashi-Maskawa (CKM)
quark mixing matrix, nucleon structure and QCD, electroweak measurements,
neutrino oscillations and, finally, studies of charmed hadrons.

  With huge samples of flavor tagged events, a MURINE should be
able to make impressive measurements of the
absolute squares of several of the elements in the
CKM quark mixing matrix.
The analyses would be analagous to, but vastly superior to,
current neutrino measurements of ${\rm |V_{cd}|^2}$ that
use dimuon events for final state tagging of charm quarks.
The current, experimentally determined values for the 9 mixing probabilities
are given in table~\ref{ckm_table}~\cite{ckm}, along with their current
percentage uncertainties and speculative projections~\cite{nufnal97} 
for how 4 of the 9 uncertainties could be reduced by a MURINE at a 500 GeV
CoM muon collider. Additionally, if muon colliders eventually reach
the 100 TeV energy scale then the associated neutrino beams will even produce
final states containing a top quark, almost certainly resulting in uniquely
precise determinations of ${\rm |V_{td}|^2}$ and ${\rm |V_{ts}|^2}$.

\begin{table}[ht!]
\caption{Absolute squares of the elements in the
Cabbibo-Kobayashi-Maskawa (CKM) quark mixing matrix.
The second row for each quark gives
current percentage uncertainties in quark mixing probabilities
and speculative projections of the uncertainties after analyses on
$10^{10}$ events from
a MURINE at a 500 GeV CoM muon collider.
The two uncertainties in brackets have not been measured directly
from tree level processes. The uncertainties assume that no unitarity
constraints have been used.}
\begin{tabular}{|c|lll|}
\hline
          & \hspace{0.2 cm} \bf{d} & \hspace{0.2 cm} \bf{s}  &
                                          \hspace{0.3 cm}\bf{b}  \\
\hline
\bf{u}    &   \bf{0.95}  &  \bf{0.05}    &  \bf{0.00001}  \\
          &   $\pm$0.1\%  &  $\pm$1.6\%    & $\pm$50\% $\rightarrow$ 1-2\% \\
          &&& \\
\bf{c}    &   \bf{0.05}  &  \bf{0.95}    &  \bf{0.002}  \\
          &   $\pm$15\% $\rightarrow$ 0.2-0.5\%   &
              $\pm$35\% $\rightarrow$ $\sim 1$\%         &
              $\pm$15\% $\rightarrow$ 3-5\%     \\
          &&& \\
\bf{t}    &  \bf{0.0001}  &  \bf{0.001}  &  \bf{1.0} \\
          &   ($\pm$25\%)   &  ($\pm$40\%)   & $\pm$30\%
\label{ckm_table}
\end{tabular}
\end{table}

  Another major motivation for MURINE's is the potential for greatly improved
measurements of nucleon structure functions (SF).
Knowledge of these SF's is crucial for
precision measurements in neutrino physics, charged lepton scattering
experiments and some precision analyses at proton-proton and lepton-proton
colliders. Further, they provide
important tests of quantum chromodynamics (QCD), and a MURINE
might well be the best single experiment of any sort for the examination of
perturbative QCD~\cite{nufnal97}.

  Neutrino physics has also had an important historical role in measuring
the electroweak mixing angle, which is simply related to the mass ratio of
the W and Z intermediate vector bosons: 
\begin{equation}
\sin^2\theta_W \equiv 1 - \left( \frac{M_W}{M_Z} \right) ^2.
                        \label{eq:wma}
\end{equation}
(To be precise, this is the Sirlin on-shell definition of $\sin^2\theta_W$.)

 Now that $M_Z$ has been precisely measured at LEP, measurements of
$\sin^2\theta_W$ in neutrino physics can be directly converted to
predictions for
the W mass. The comparison of this prediction with direct $M_W$
measurements in collider experiments constitutes a precise prediction
of the SM and a sensitive test for exotic physics modifications to
the SM~\cite{deltamw}.
Reference~\cite{nufnal97} estimates that the
predicted uncertainty in $M_W$ from a MURINE analysis might be of order
10 MeV, which improves by an order of magnitude on
today's neutrino experiments~\cite{deltamw,ccfrwma} and is
approximately equal to the projected best direct measurements from future
collider experiments.

  There are currently several experimental indications~\cite{nuosc} that
neutrinos might have non-zero masses and oscillate in flight between the
flavor eigenstates. The probability for an oscillation
between two of the flavors is given by\cite{pdg}:
\begin{equation}
{\rm Oscillation\; Probability = \sin ^2 \theta \times
       \sin ^2 \left( 1.27 \frac{\Delta m^2 [{\rm eV^2}].L[km]}
                                            {E_\nu [GeV]} \right),  }
                                        \label{eq:oscprob}
\end{equation}
where the first term gives the mixing strength and the second term gives the
distance dependence. Reference~\cite{nufnal97} obtains
the following order-of-magnitude mass limit for an assumed long-baseline
detector with reasonable parameters and with full mixing:
\begin{equation}
{\rm \Delta m^2 |_{min} \sim O(10^{-4})\: eV^2, }
                                        \label{eq:deltamsq}
\end{equation}
relatively independent of the distance to the detector.
Similarly, a mixing probability sensitivity
for $10^{10}$ events in a short-baseline detector is found to be as low as
\begin{equation}
{\rm \sin ^2 \theta |_{min} \sim O(10^{-7}), }
                                        \label{eq:thetaosc}
\end{equation}
for the most favorable value of $\Delta m^2$.
Both of these estimates apply generically to all
3 possible mixings between 2 flavors:
$\nu_e \leftrightarrow \nu_\mu$,
$\nu_e \leftrightarrow \nu_\tau$ and
$\nu_\mu \leftrightarrow \nu_\tau$.
(See also reference~\cite{Geer} for another discussion of neutrino
oscillations at a MURINE.)

The $\Delta m^2$ estimate is more than an order
of magnitude better than any proposed accelerator or reactor experiments for
$\nu_\mu \leftrightarrow \nu_\tau$
and $\nu_e \leftrightarrow \nu_\tau$, and is competitive with the best such
proposed experiments for $\nu_e \leftrightarrow \nu_\mu$.
The estimated value for $\sin ^2 \theta |_{min}$ is even more impressive
-- orders of magnitude better than in any other current or
proposed experiment for each of the three possible oscillations.
Such an experiment would either convincingly refute or accurately
characterize the claimed observations of oscillations by both the
LSND and Super-Kamiokande collaborations.

  As an interesting final topic, MURINE's should be rather impressive
factories for the study of charm -- with
a clean, well reconstructed sample of several times $10^8$ charmed hadrons
produced in $10^{10}$ neutrino interactions.
There are several interesting physics motivations for charm
studies at a MURINE~\cite{charmphysics}. As an example, particle-antiparticle
mixing has yet to be observed in the charm sector~\cite{d0mixing}, and 
it is quite plausible~\cite{nufnal97}
that a MURINE would provide the first observation of
${\rm D^0 - \overline{D^0}}$ mixing.

\section*{Muon Collider Scenarios}
%%%%%%%%%%%%%%%%%%%%%%%%%%%%%%%%%%
\label{sec-scenarios}

  This section explores the collider physics opportunities at muon
colliders through reference to the example collider parameter sets of
table~\ref{specs}. Complementary discussion on the collider physics
aspects of these parameters can be found in~\cite{epac98}, where it
is opined that each parameter set has some aspects that appear
challenging but none of the parameter sets are obviously implausible.
Admittedly, table 1 gives a rather incomplete sampling of the
possibilities and, for example, discussions of additional physics
options with sub-TeV muon colliders may be found in~\cite{feworkshop}.

 The section begins with a discussion on muon collider design
constraints due to the potential neutrino radiation hazard -- a serious
problem that is unique to muon colliders -- before examining, in turn,
the physics potential of each of the parameter sets in
table~\ref{specs}.

\subsection*{The Potential Radiation Hazard from Neutrinos}
%%%%%%%%%%%%%%%%%%%%%%%%%%%%%%%%%%%%%%%%%%%%%%%%%%%%%%%%%%%
\label{subsec-radhazard}

  A serious and unexpected problem that has arisen for multi-TeV \mumu
colliders is the potential radiation hazard posed by neutrinos emitted from
muon decays in the collider ring~\cite{phdnote,bjkrad}. These
neutrinos produce a
``radiation disk'' in the plane of the ring, and the potential radiation
hazard results from the showers of ionizing particles from occasional
neutrino interactions in the soil and other objects bathed by this disk.
Although the neutrino cross-section is tiny, this is greatly
compensated by the enormous number of tightly collimated high energy
neutrinos produced at the collider ring.

  With some reasonable assumptions, the approximate average numerical
value for the annual radiation dose in the plane of the collider ring is
easily derived to be~\cite{bjkrad}:
\begin{equation}
{\rm D_{ave}[mSv/yr]} \simeq
      {\rm 0.044 \times \frac{ f_b[Hz] \times N_0[10^{12}]
                 \times (1-e^{-t_D[\gamma \tau_\mu]})
                 \times (E_\mu[TeV])^3 }{D[m]} }
                                  \label{eq:nurad},
\end{equation}
with notation as in table~\ref{specs} and assuming an accelerator running
time of $10^7$ seconds per year. For comparison, the U.S.
federal off-site radiation limit is 1 mSv/year, which is of the same order of
magnitude as the typical background radiation from natural causes (i.e.
0.4 to 4 mSv/yr~\cite{pdg}).

 To explain the form of equation~\ref{eq:nurad}, the inverse
dependence of the neutrino radiation
on the collider depth arises because the radiation levels fall
as the inverse square of the distance from the ring while the distance
to reach the Earth's surface, assuming a spherical Earth, goes as the square
root of the depth. Also, the cubic dependence on the collider energy
comes from combining the approximately linear rises with energy of a) the
neutrino cross section b) the energy deposited per interaction, and c) the
beam intensity due to the decreasing angular divergence of the neutrinos
in the vertical plane(equation~\ref{eq:thetanu}). (There are actually some
mitigating factors that come into play at the highest energies and are not
included in equation~\ref{eq:nurad}~\cite{bjkrad}.)

 This equation is not intended to be accurate at much better
than an order of magnitude level and is deliberately conservative,
i.e. it may well overestimate the radiation levels. Because of the
energy dependence, the radiation levels rapidly become a serious
design constraint for colliders at the TeV scale and above.  

  The radiation intensity may be greatly enhanced downstream from
straight sections in the collider ring, with the additional intensity
rising in proportion to the length
of the straight section. As a benchmark,
the length of straight section to produce ten times the
planar average dose, ${\rm L_{x10}}$, may be shown~\cite{bjkrad} to be
approximately:
\begin{equation}
{\rm L_{x10} [meters] \simeq 0.3 \times \frac{C[km]}{E_\mu[TeV]} }
                                    \label{eq:Lx10}.
\end{equation}
This equation shows that the intensity from the straight section picks up
another power of the collider energy, which is due to the falling
horizontal angular divergence, but this is approximately compensated for
by the collider circumference also rising in approximate
proportion to the beam energy.
As can be seen from table~\ref{specs}, ${\rm L_{x10}}$ is only of order
a meter at all collider energies, so great care must be taken in the
design of the collider ring to minimize or eliminate long straight
sections.

  Because of the cubic rise with energy of the neutrino radiation intensity,
muon colliders at CoM energies of beyond a few TeV will
probably have to be constructed
at isolated sites where the public would not be exposed to
the neutrino radiation disk. Such sites clearly exist, perhaps even with
useful existing infrastructure. (An extreme example would be close to a
nuclear test site, such as in Nevada, U.S.A.) These will presumably be
``second generation'' machines, arriving after the technology of muon
colliders has been established in one or more smaller and less expensive
machines built at existing HEP laboratories.

\subsection*{An S-Channel Higgs Factory}
\label{subsec-higgs}

  Besides exploring the physics at the energy frontier, muon colliders
with very narrow CoM energy spreads
are particularly suited to both resonance production
and threshold studies of elementary particles.
The principal example of such a resonant process is the s-channel
production of Higgs bosons. The relatively strong coupling strength of
muons to the Higgs channel -- approximately 40 000 times that for electrons
-- gives \mumu colliders a unique potential to study this process.

 The first parameter set in table 1 is intended for precision studies
of a 100 GeV SM-like Higgs boson, hypothesized to have been discovered
previously at either LEP, the Tevatron or the LHC. (Of course, the
CoM energy of the collider would actually be fixed at the true Higgs mass.)
The low CoM energy spread has been chosen to reproduce the
predicted width of a SM Higgs at this energy: 2 to 3 MeV. After an initial
coarse scan
to find the exact energy of the resonance, a fine scan of the resonance
would provide uniquely precise measurements of the Higgs mass, width
and cross-section.

 The technological issues specific to these Higgs factory parameters
have been evaluated in some detail over the past year by the
MCC~\cite{status}. For example, a collider magnet
lattice has been designed for the narrow momentum spread beam and
the required precise beam calibration was found to be possible by measuring
the rate of the muon spin precession.

 The physics case for an s-channel Higgs factory has also been studied in
some detail~\cite{feworkshop}. The effectiveness of the collider obviously
depends on the existence of a Higgs boson in the appropriate mass range.
If the Higgs is too light then, at currently assumed luminosities, the signal
will be buried in the backgrounds from the Z resonance. On the other hand,
the Higgs width increases with mass, becoming too broad for effective
study beyond about 150 GeV. The following approximate scenarios emerge
for a SM or SM-like (e.g. supersymmetric) Higgs:
\begin{enumerate}
\item ${\rm M_H < 105\; GeV}$: probable discovery at LEP. Backgrounds
probably too high for an s-channel Higgs factory.
\item ${\rm 105\; GeV < M_H < 150\; GeV}$: fairly likely to be discovered
at FNAL but with poor mass resolution. An s-channel Higgs factory will become
useful following more precise ${\rm M_H}$ measurements from the LHC and/or
a future lepton collider with a CoM energy of a few hundred GeV.
\item ${\rm M_H > 150\; GeV}$ (this is now experimentally disfavored):
the resonance would be too broad for an s-channel factory.
\item (for completeness) no Higgs. A Higgs factory is obviously not useful.
\end{enumerate}

  As an example of a detailed study for the Higgs mass in a favorable
region, reference~\cite{feworkshop} makes predictions
for ${\rm 0.4\; fb^{-1}}$ of on-peak data and a SM-like
Higgs with ${\rm M_H = 110\; GeV}$. They predict a resolution of
approximately 0.1 MeV on the Higgs mass, 0.5 MeV on the width 
and branching ratio determinations as accurate as 3 percent (for
the ${\rm H \rightarrow b \overline{b}}$ channel).
These precise measurements on such an important elementary particle
clearly provide strong motivation for this muon collider option,
either as a stand-alone ``first muon collider'' or as a relatively
inexpensive add-on to a complex with a higher energy machine.

\subsection*{A 1 TeV Muon Collider to Complement the LHC}
%--------------------------------------------------------
\label{subsec-1tev}

 The motivation for a 1 TeV muon collider would be roughly the same as that
of proposed \ee linear colliders at the 1 TeV energy scale -- that is, to
perform precision studies on whatever elementary particles are discovered at
the LHC hadron collider and to search for new particles that will not be
evident in the physical and experimental conditions of the LHC.
Thus, a 1 TeV muon collider may be considered as a valuable back-up
technology in case electron colliders at this energy either run into unforeseen
technical difficulties or are found to be unacceptably expensive.
Further, such a muon collider may have a role to play even if a 1 TeV e+e-
collider is built, due to potentially different physics processes (e.g.
Higgs-type particle production) and also to differences in the beam
specifications, as follows.

  One TeV electron colliders should
be able to achieve higher levels of polarization than their muon collider
counterparts, which may have polarization levels in the region of
20\%~\cite{book}. On the other hand, beamstrahlung at 1 TeV electron
colliders will result in roughly a 10\% fractional
spread in collision energy rather than the parts-per-mil spreads
assumed for muon colliders. Thus, electron colliders will be favored
for studies where high polarization is important while muon colliders
should do better in studies of resonances and in other processes where the
CoM energy constraint is important.

%%% start of EPAC text

  The 1 TeV parameter set of table 1 would give about the same luminosity as,
for example, the design for the proposed NLC linear electron
collider~\cite{NLC} at
the same energy, and the physics motivation and capabilities might be
relatively similar. Placement of the collider at 125 meters depth
(the approximate depth of the existing LEP/LHC tunnel at CERN)
reduces the average neutrino radiation in the collider plane to
less than one thousandth of the U.S. federal off-site radiation limit.
(As already mentioned, attention would still need to be paid to minimizing
the length of any low-divergence straight sections in the collider ring.)

%%% end of EPAC text

\subsection*{A Muon Collider at the Energy Frontier: 4 TeV}
%----------------------------------------------------------
\label{subsec-4tev}

 Muon colliders appear to have much greater potential than electron
colliders to push to lepton collision energies above the LHC mass
reach (which might be roughly 1 to 2 TeV, depending on the process).
The 4 TeV parameter set was chosen as being at about the highest energy
that is practical for a ``first generation'' muon collider on an existing
laboratory site, due to neutrino radiation.

 The same comments about neutrino radiation apply as in the 1 TeV design
and, in addition, it is necessary to greatly reduce the muon current,
accepting the consequent loss in luminosity. The assumed 300 meter depth
happens to correspond approximately to appropriate bedrock formations at
both the BNL and Fermilab HEP laboratories.

 Even the reduced luminosity of this parameter set, 
$6.2 \times 10^{33} {\rm cm^{-2}.s^{-1}}$, appears sufficient to discover
whatever elementary particles lie in the mass range of 1 to 4 TeV (i.e. beyond
the reach of the LHC), provided only that the experimental signature for
production is not particularly obscure and the production cross-section
is not greatly suppressed relative to typical SM couplings (as exemplified
by the benchmark process, $\mu\mu \rightarrow {\rm ee}$).

 An added motivation for building a ``first generation'' muon collider
at the highest possible energy is that this would provide the best
technical foundation for construction of the very high energy,
high luminosity muon colliders (10 TeV and above) that are the ultimate
goal of muon collider technology.

\subsection*{A Second Generation Muon Collider at 10 TeV}
%--------------------------------------------------------
\label{subsec-10tev}

  The 4th parameter set of table~\ref{specs} specifies a ``second
generation'' muon collider at 10 TeV CoM energy, assumed to be constructed
at a site where neutrino radiation is not a constraint (see the previous
subsection on neutrino radiation). It is seen that the relaxed neutrino
radiation constraint might allow an exciting luminosity of
$1.0 \times 10^{36} {\rm cm^{-2}.s^{-1}}$ at several times the
discovery mass reach of the LHC, making this collider an exciting
prospect for the future progress of HEP.

  Besides mapping out the spectrum of elementary particles in the energy
decade up to 10 TeV, it is further reasonable to assume that anything
already discovered at the LHC could be more fully studied in the much cleaner
physics environment of such a lepton collider. Particles in the 100 GeV to
1 TeV range should be copiously produced through higher order processes
in a 10 TeV muon collider, as evidenced by the production, via the
WW-fusion process, of order ten million SM Higgs particles per year
(assuming it exists with a mass below 1 TeV).

 The technical difficulties specific to muon colliders at this energy scale and
above have yet to be assessed in detail. It is comforting that relativistic
kinematics makes the acceleration of the muons progressively easier at higher
energies due to a rising muon lifetime, shrinking transverse bunch size
and reduced sensitivity to disruptive influences such as wake fields.
On the other hand, detector backgrounds involving high energy muons will
clearly become more challenging, as will the design and layout of the
final focus magnets around the ip (see ~\cite{epac98} for details).
To put this in perspective, these technical challenges will need to be
compared with the considerable challenges that are essentially independent
of the collider energy -- particularly the construction and operation
of the muon cooling channel.

\subsection*{The Ultimate Energy Scale for Muon Colliders: 100 TeV}
%------------------------------------------------------------------
\label{subsec-100tev}

 The highest energy parameter set in table 1, at 100 TeV, represents
what is likely the ultimate energy scale for muon colliders, with
a mass reach for discovering elementary particles that
is probably inaccessible even to hadron colliders.

 The parameter set assumes technical extrapolations beyond today's
limits and presents easily the most difficult design challenge
of all the parameter sets, for the following reasons:
\begin{itemize}
 \item  cost reductions will be needed to make a machine of this size
        affordable
 \item  siting will be more difficult than at 10 TeV, since the neutrino
        radiation is now well above the U.S. federal limit
 \item  the final focus design is much more difficult even than at 10 TeV,
        as illustrated by the much larger demagnification factor
        (see \cite{epac98} for further discussion)
 \item  the muon bunches, although much smaller than in the other sets, are
        also much cooler (again, see \cite{epac98} for further discussion)
 \item  the beam power has risen to 170 MW, with synchrotron radiation
        rising rapidly to contribute a further 110 MW.
\end{itemize}
It seems reasonable to assume that the rapid rise to prominence of the
synchrotron radiation will effectively prohibit muon colliders at the PeV
energy scale, even if the other challenges could be negotiated.

 A 100 TeV muon collider is clearly not a near-term prospect. However, the
unique opportunity to explore the physics at this energy scale could well
turn out to be crucial in unlocking the profound mysteries of the elementary
particle spectrum  and its role in the universe. With such compelling
motivation, it is certainly not ruled out that a muon collider at this energy
scale could become achievable after a couple of decades of dedicated research
and development.

\section*{Summary}
%%%%%%%%%%%%%%%%%%
\label{sec-summary}

 An overview has been given of the potential prospects for neutrino physics
and collider physics at muon colliders and it has been shown that muon
colliders may well come to assume a central role in the future of experimental
high energy physics. Their discovery reach for new elementary particles
might eventually be in the region of 100 TeV and they could also open up
exciting new vistas in neutrino physics and other precision studies.

 This provides strong motivation for a
continuing and expanding vigorous research and development program in 
muon collider technology, and such a program will be needed to make
muon colliders a reality on an attractive timescale.

\end{document}